\documentclass[twocolumn]{aastex7}
\newcommand{\commentred}[1]{#1}
\newcommand{\commentr}[1]{#1}

\newcommand{\commentbf}[1]{#1}

\usepackage{ulem}
\usepackage{comment}
\usepackage{xcolor}
\DeclareRobustCommand{\erase}{\bgroup\markoverwith{\textcolor{red}{\rule[.5ex]{2pt}{0.4pt}}}\ULon}

\begin{document}

\title{Multiphase Gas Structure in the Circumnuclear Region of NGC 5506 Observed with ALMA}

\shorttitle{Multiphase Gas Structure in the CND of NGC 5506}
\shortauthors{Takechi et al.}

\correspondingauthor{Kana Takechi}
\email{c251435s@alm.icu.ac.jp}

\author{Kana Takechi}
\affiliation{International Christian University, 3-10-2 Osawa, Mitaka, Tokyo 181-8585, Japan}
\email{c251435s@alm.icu.ac.jp}

\author[0000-0003-0292-3645]{Hiroshi Nagai}
\affiliation{National Astronomical Observatory of Japan, 2-21-1 Osawa, Mitaka, Tokyo 181-8588, Japan}
\affiliation{Department of Astronomical Science, The Graduate University for Advanced Studies, SOKENDAI, 2-21-1 Osawa, Mitaka, Tokyo 181-8588, Japan}
\email{hiroshi.nagai@nao.ac.jp}

\author[0000-0003-2535-5513]{Nozomu Kawakatu}
\affiliation{National Institute of Technology, Kure College, 2-2-11, Agaminami, Kure, Hiroshima 737-8506, Japan}
\email{kawakatsu@kure-nct.ac.jp}

\author[0000-0002-8779-8486]{Keiichi Wada}
\affiliation{Kagoshima University, 1-21-35, Korimoto, Kagoshima 890-0065, Japan}
\email{wada@astrophysics.jp}

\author[0000-0001-9452-0813]{Takuma Izumi}
\affiliation{National Astronomical Observatory of Japan, 2-21-1 Osawa, Mitaka, Tokyo 181-8588, Japan}
\affiliation{Department of Astronomical Science, The Graduate University for Advanced Studies, SOKENDAI, 2-21-1 Osawa, Mitaka, Tokyo 181-8588, Japan}
\email{takuma.izumi@nao.ac.jp}

\author[0000-0002-2709-7338
]{Motoki Kino}
\affiliation{Kogakuin University of Technology \& Engineering, Academic Support Center, 2665-1 Nakano-machi, Hachioji, Tokyo 192-0015, Japan}
\affiliation{National Astronomical Observatory of Japan, 2-21-1 Osawa, Mitaka, Tokyo 181-8588, Japan}
\email{motoki.kino@gmail.com}

\author[0000-0002-6939-0372
]{Kouichiro Nakanishi}
\affiliation{National Astronomical Observatory of Japan, 2-21-1 Osawa, Mitaka, Tokyo 181-8588, Japan}
\affiliation{Department of Astronomical Science, The Graduate University for Advanced Studies, SOKENDAI, 2-21-1 Osawa, Mitaka, Tokyo 181-8588, Japan}
\email{nakanisi.k@nao.ac.jp}

\author{Naoki Isobe}
\affiliation{Institute of Space and Astronautical Science (ISAS), Japan Aerospace Exploration Agency (JAXA), 3-1-1 Yoshinodai, Chuo-ku, Sagamihara, Kanagawa 252-5210,
Japan}
\email{n-isobe@ir.isas.jaxa.jp}

\author[0000-0001-5946-9960]{Mahito Sasada}
\affiliation{Department of Physics, Tokyo Institute of Technology, 2-12-1
Ookayama, Meguro-ku, Tokyo 152-8551, Japan}
\email{sasada.m.ab@m.titech.ac.jp}

\author[0000-0003-4384-9568
]{Akihiro Doi}
\affiliation{Institute of Space and Astronautical Science (ISAS), Japan Aerospace Exploration Agency (JAXA), 3-1-1 Yoshinodai, Chuo-ku, Sagamihara, Kanagawa 252-5210,
Japan}
\affiliation{Department of Space and Astronautical Science, SOKENDAI, 3-1-1 Yoshinodai, Chuou-ku, Sagamihara, Kanagawa \\252-5210, Japan}
\email{doi.akihiro@jaxa.jp}

\begin{abstract}

We present a study of the multiphase gas structure and kinematics of the circumnuclear disk (CND) of NGC 5506, a nearby edge-on Seyfert galaxy, at a spatial resolution of $\sim20$~pc. Observations of [\ion{C}{1}](1--0), CO(3--2), and HCO$^{+}$(4--3) obtained with the Atacama Large Millimeter/submillimeter Array reveal the CND dominated by rotational motion on scales of several hundred parsecs. No significant differences in geometrical thickness or velocity structure are found between [\ion{C}{1}](1--0) and CO(3--2) across the CND, whereas HCO$^{+}$(4--3) emission is more concentrated toward the disk plane. The ratio of velocity dispersion to rotational velocity, a proxy for disk scale height-to-radius ratio, is high ($\gtrsim0.9$) in the central region ($\lesssim30$~pc) for both [\ion{C}{1}](1--0) and CO(3--2), indicating geometrically thick structures in both tracers.
Regions where the [\ion{C}{1}](1--0)/CO(3--2) ratio exceeds the CND average are spatially correlated with the [\ion{O}{3}]$\lambda5007$ bicone observed with the Hubble Space Telescope, suggesting that CO is preferentially dissociated by the AGN-driven biconical ionized outflow. 
The observed CND scale height and velocity dispersions traced by [\ion{C}{1}](1--0) and CO(3--2) are consistent with a model in which supernova-driven turbulence provides the vertical support for the CND.

\end{abstract}


\section{Introduction} \label{sec:intro}
An active galactic nucleus (AGN) is the compact central region of a galaxy characterized by intense electromagnetic radiation.
In the AGN unified model, 
the presence or absence of broad emission lines is attributed to the viewing angle relative to a geometrically and optically thick torus-shaped structure~\citep[hereafter torus;][]{Antonucci1993, Urry1995}. 
Constraining the physical properties of the torus is essential for understanding supermassive black hole (SMBH)--host galaxy coevolution~\citep[e.g.,][]{Kormendy2013}, as this structure is thought to play a crucial role in the mass supply from the host galaxy to the SMBH. 

The torus has been investigated both theoretically and observationally.
Early continuous dust torus models failed to reproduce the observed 9.7~$\mathrm{\mu m}$ silicate feature~\citep{Pier1992,Pier1993}. This motivated the development of clumpy dust models~\citep[e.g.,][]{Nenkova2002, Nenkova2008a, Nenkova2008b, Elitzur2006b, Honig2006, Schartmann2008, Schartmann2009, Levenson2009}. 
The high observed fraction of type-2 AGN implies that the torus is geometrically thick~\citep[e.g.,][]{Lawrence2010, Merloni2014, Davies2015}. However, the mechanism responsible for supporting this geometrical thickness remains uncertain. 
\citet{Krolik2007} presented an axisymmetric torus model supported by infrared radiation pressure~\citep[see also][]{Shi2008}. 
Several models have incorporated the effects of nuclear starburst activity, which is commonly observed in AGN host galaxies~\citep[e.g.,][]{Heckman1997, Cid2001, Gonzalez2001, Davies2007}. \citet{Ohsuga1999, Ohsuga2001} proposed that radiation pressure from both the AGN and the nuclear starburst can support a dusty obscuring wall.
A supernova (SN)-driven turbulent torus model was introduced by \citet[][]{Wada2002}. 
\citet{Kawakatu2020} subsequently investigated the dependence of the AGN obscuring fraction on black hole mass and AGN luminosity, incorporating both SN-driven turbulence and AGN radiative feedback.

Polar elongation observed in the mid-infrared \citep[e.g.,][]{Tristram2014, Asmus2016} suggests that the torus is dynamically formed rather than static. The radiation-driven fountain model~\citep{Wada2012} was proposed to explain these observations. Three-dimensional hydrodynamic simulations based on this framework demonstrate that the torus thickness can be dynamically maintained through gas circulation~\citep{Wada2012}. The model was extended to include SN feedback and nonequilibrium chemistry in X-ray-dominated regions~\citep{Wada2016}, allowing detailed predictions for the distributions of atomic, molecular, and ionized gas. This model has been tested against multiwavelength observations of the Circinus galaxy, a nearby Seyfert galaxy~\citep{Izumi2018, Wada2018a, Wada2018b, Uzuo2021, Tanimoto2023, Baba2024}. In particular, \citet{Izumi2018} identified kinematic signatures in [\ion{C}{1}](1--0) and CO(3--2) consistent with the model predictions, although these were not evident in subsequent higher-resolution observations~\citep{Izumi2023}.

The development of high angular resolution facilities such as the Atacama Large Millimeter/submillimeter Array (ALMA) has enabled observations of AGN circumnuclear regions on scales of a few to several tens of parsecs~\citep[e.g.,][]{Imanishi2018, Combes2019, Garcia2021,Izumi2023}. 
Nevertheless, the physical properties and formation mechanisms of the torus remain poorly constrained, motivating high-resolution observations of additional nearby systems. 
In this study, we analyze ALMA and Hubble Space Telescope (HST) observations of the circumnuclear region of NGC~5506 to characterize the distribution and kinematics of atomic, molecular, and ionized gas, and to compare these results with theoretical models.

NGC~5506 is a nearby Seyfert galaxy classified in the literature as type-2~\citep{Trippe2010}, type-1.9~\citep{Maiolino1995}, or narrow-line Seyfert 1~\citep{Nagar2002}. The circumnuclear disk (CND) of NGC~5506 is viewed at a nearly edge-on inclination, which enhances the sensitivity to differences in geometrical thickness among multiple gas phases. Throughout this paper, we define the CND of NGC~5506 as the region where emission is detected at $>3.5\sigma$ within a 5$^{\prime\prime}\times1^{\prime\prime}.4$ box centered on the AGN. 
NGC~5506 hosts an SMBH with a mass of $2.0^{+8.0}_{-1.6}\times 10^7~M_{\odot}$ \citep{Gofford2015}, derived as the logarithmic mean of two independent estimates: $8.8\times10^{7}~M_{\odot}$ \citep{Papadakis2004}, and $5.11\times10^{6}~M_{\odot}$ \citep{Nikolajuk2009}. 
The bolometric luminosity is 1.3~$\times$~10$^{44}$~erg~s$^{-1}$~\citep{Davies2020}, and the Eddington ratio is $\lambda_{\rm Edd} = 0.05^{+0.21}_{-0.04}$ \citep{Esposito2024}.
Outflow activity in NGC~5506 has been confirmed by the detection of an ultrafast outflow~\citep{Gofford2013,Gofford2015} and radio jet emission \citep{Kinney2000,Roy2000, Roy2001}. 
HST imaging and kinematic modeling of [\ion{O}{3}]$\lambda5007$ reveal a biconical ionized outflow with an inclination of $10^\circ$ and a maximum velocity of 550~km~s$^{-1}$ on scales of 300~pc~\citep{Fischer2013}. We assume a flat $\Lambda$CDM cosmology with $H_0=67.66$~km~s$^{-1}$~Mpc$^{-1}$, $\Omega_{\rm m}=0.3111$, and $\Omega_{\Lambda}=0.6889$~\citep{Planck2020}. 
We adopt a luminosity distance of 23.8~Mpc for NGC~5506~\citep{Karachensev2014}, at which 1$^{\prime\prime}$ corresponds to 114~pc.

\section{Observation and Data Analysis} \label{sec:data}

\begin{table*}
\begin{center}
    \caption{Properties of Emission Lines}
    \begin{tabular}{ccccccc}
    \hline\hline
         Line Name & Species & Transition & Wavelength & Frequency & Critical Density & Reference$^{1}$\\ 
         & & & ($\mathrm{\mu m}$) & (GHz) & (cm$^{-3}$)& \\
         \hline
        $[$\ion{C}{1}$]$(1--0) & C & $^3P_1\rightarrow^3P_0$ & 370.4~$\rm$ & 492.2 & $4.7\times10^2$ & (1) \\
        CO(3--2) & CO & $J=3\rightarrow2$ & 867.0 & 345.8 & $8.4\times10^3$ & (2)\\
        HCO$^{+}$(4--3) & HCO$^{+}$ & $J=4\rightarrow3$ & 840.4
 & 356.7 & $1.8\times10^6$ & (2)\\
         $[$\ion{O}{3}$]$$\lambda5007$ & $\rm O^{\commentred{2+}}$ & $^1D_2\rightarrow^3P_2$ & $500.7\times10^{-3}$ & $598.7\times10^3$  & $6.8\times10^5$ & (3)\\
        \hline\hline
    \end{tabular}
    \label{tbl:lines}
\end{center}
\tablecomments{$^1$References of the critical densities: (1) \cite{Carilli2013}; (2) \cite{Greve2009}; and (3) \cite{Osterbrock2006} }
\end{table*}

\begin{table*}
\begin{center}
\caption{Observational Parameters \label{tbl2}}
\begin{tabular}{ccccccc}
\hline\hline
Line & Project ID & Date$^1$ & $\#$ Antennas$^2$ & Baseline$^3$ & On-source Time$^4$ & Pipeline Version$^5$ \\ 
 & & (UT) & & (m) & (h:mm:ss) & \\
\tableline
[\ion{C}{1}](1--0) & 2022.1.00410.S& 2023 Apr 13, 20 & 42 & 14--1198 & 2:00:57 & 2022.2.0.68 \\
CO(3--2) & 2017.1.00082.S & 2017 Dec 31 & 44 & 14--2309 & 0:33:24 &Pipeline-CASA51-P2-B \\ 
HCO$^{+}$(4--3) & 2017.1.00082.S & 2017 Dec 31 & 44 & 14--2309 & 0:33:24 &Pipeline-CASA51-P2-B\\

\hline
\hline
\end{tabular}
\end{center}
\tablecomments{$^1$Date(s) of observation. $^2$Number of antennas. $^3$Shortest and longest baseline length. $^4$Total on-source integration time. $^5$CASA pipeline version used for data reduction.}
\end{table*}

\begin{table*}
\begin{center}
\caption{Imaging Properties \label{tbl1}}
\begin{tabular}{ccccccc}
\hline\hline
Emission & Telescope & $\nu_{\rm rest}$ & Beamsize, PA & Beamsize & rms & Peak \\ 
 & & (GHz) & (arcsec $\times$ arcsec), (deg)  & (pc $\times$ pc) & (mJy beam$^{-1}$) & (mJy beam$^{-1}$) \\
\tableline
[\ion{C}{1}](1--0)$^1$ & ALMA & 492.2 & 0.166 $\times$ 0.143, (\commentred{70.6}) & 19.1 $\times$ 16.4 & 7.7 & 87 \\
CO(3--2)$^2$ & ALMA & 345.8 & 0.176 $\times$ 0.135, (\commentred{-52.8}) & 20.2 $\times$ 15.5 & 2.2 & 56 \\ 
HCO$^{+}$(4--3)$^2$ & ALMA & 356.7 & 0.174 $\times$ 0.134, (\commentred{-54.4}) & 20.0 $\times$ 15.4 & 0.69 & 4.7\\
$[$\ion{O}{3}$]$ $\lambda5007^3$ & HST & $598.7\times10^{3}$ & ... & ... & ... & ... \\ 
\hline
\hline
\end{tabular}
\end{center}
\tablecomments{${}^1$Project ID: 2022.1.00410.S (PI: H. Nagai); ${}^2$Project ID: 2017.1.00082.S \citep[PI. S. Garc{\'\i}a-Burillo;][]{Garcia2021}; ${}^3$Observation ID: x2740302t (PI: F. D. Macchetto)}
\end{table*}

We analyze ALMA and HST observations of NGC~5506 to investigate the multiphase gas structure and kinematics of the CND. The emission lines used in this study are [\ion{C}{1}](${}^3P_1-{}^3P_0$; hereafter 1--0), CO($J=3\textrm{--}2$; hereafter 3--2), HCO$^{+}$($J=4\textrm{--}3$; hereafter 4--3), and [\ion{O}{3}]$\lambda$5007, which trace atomic, molecular, dense molecular, and ionized gas, respectively. Line properties are listed in Table \ref{tbl:lines}. The [\ion{O}{3}]$\lambda$5007 line was observed with HST (observation ID: x2740302t); the other lines were observed with ALMA in Cycle 5 (project ID: 2017.1.00082.S) and Cycle 9 (project ID: 2022.1.00410.S). Observational parameters are summarized in Table \ref{tbl2}.

\subsection{ALMA}
We observed [\ion{C}{1}](1--0) with ALMA Band 8 in Cycle 9. 
CO(3--2) and HCO$^{+}$(4--3) were obtained from archival Band 7 observations. 
Baseline lengths are 14--1198~m for [\ion{C}{1}](1--0) and 14--2309~m for CO(3--2) and HCO$^{+}$(4--3); the ranges of $\it{uv}$-distance in units of wavelength are comparable across all three datasets. 
Calibration was performed using the ALMA pipeline in Common Astronomy Software Applications~\citep[CASA, versions listed in Table~\ref{tbl2};][]{Casa2022}, with one, two, and two antennas flagged for [\ion{C}{1}](1--0), CO(3--2), and HCO$^{+}$(4--3), respectively. 
Calibrated data were imaged using CASA (version 6.5.6) task \texttt{tclean} with a channel width of 20~km~s$^{-1}$. 
Automasking was applied following the parameters recommended in the CASA automasking guide. 
We set the \texttt{nsigma} parameter to 2, corresponding to a cleaning threshold of $2\times$ the rms noise~\citep{Casa2022}. 
Because this study compares geometrical thicknesses among different gas phases, the Briggs \texttt{robust} parameter was adjusted to synthesize comparable beam sizes across all three lines. HCO$^{+}$(4--3) data were imaged with \texttt{robust} = 2.0 (natural weighting), resulting in a synthesized beam size of $0.^{\prime\prime}174\times0.^{\prime\prime}134$. 
For [\ion{C}{1}](1--0) and CO(3--2), \texttt{robust} values of 0.8 and 1.3 were used, producing synthesized beam sizes of $0.^{\prime\prime}166 \times 0.^{\prime\prime}143$ and $0.^{\prime\prime}176 \times 0.^{\prime\prime}135$, respectively. Imaging properties are summarized in Table~\ref{tbl1}.
The ALMA Cycle 9 Proposer's Guide indicates absolute flux uncertainties of $\sim10\%$ for both Band 7 and 8. Throughout this paper, error values represent only statistical uncertainties unless otherwise noted. 

\subsection{HST}

[\ion{O}{3}]$\lambda$5007 was observed with the HST Faint Object Camera (FOC) using the f/96 aperture and the F501N filter~\citep[$\lambda_0=501$~nm and $\Delta\lambda_0=7.4$~nm;][]{Nota1996}. 
The dataset covers only the northern side of the emission~\citep[][]{Ruiz2005}.
The FOC image exhibits an astrometric offset of approximately 3$\arcsec$.
To correct the WCS coordinates, we used an HST STIS image and two reference stars within the field of view: Star A at (R.A., decl.) = (14$^{\rm h}$13$^{\rm m}$15$^{\rm s}$.91, $-$3$\arcdeg$12$\arcmin$21$\arcsec$.07) and Star B at (R.A., Decl.) = (14$^{\rm h}$13$^{\rm m}$14$^{\rm s}$.82, $-$3$\arcdeg$12$\arcmin$22$\arcsec$.42). Star A is listed in the Pan-STARRS1 (PS1) catalog and serves as the primary astrometric reference \citep{Chambers2016}, but is not detected in the FOC image. Star B is detected in the FOC image but not listed in the PS1 catalog. 
The STIS image was used to transfer the astrometric reference between Stars A and B. 
Centroids of both stars were measured using \texttt{DAOStarFinder} from the \texttt{photutils} package \citep{Bradley2024}. 
The STIS image was first aligned to the PS1 catalog, then used to determine the position of Star B. The FOC image was subsequently adjusted such that Star B coincides with this position, resulting in an astrometric accuracy better than 0$\arcsec$.2.
The point spread function (PSF) of the [\ion{O}{3}]$\lambda$5007 image was measured using \texttt{SExtractor} \citep{Bertin1996} by fitting a point source at (R.A., Decl.) = (14$^{\rm h}$13$^{\rm m}$14$^{\rm s}$.69, $-$3$\arcdeg$12$\arcmin$24$\arcsec$.90). This yields a full width at half-maximum (FWHM) of 0$^{\prime\prime}$.054.

\section{Results} \label{sec:results}

\begin{figure*}[ht!]
\begin{center}
\includegraphics[width=\linewidth]{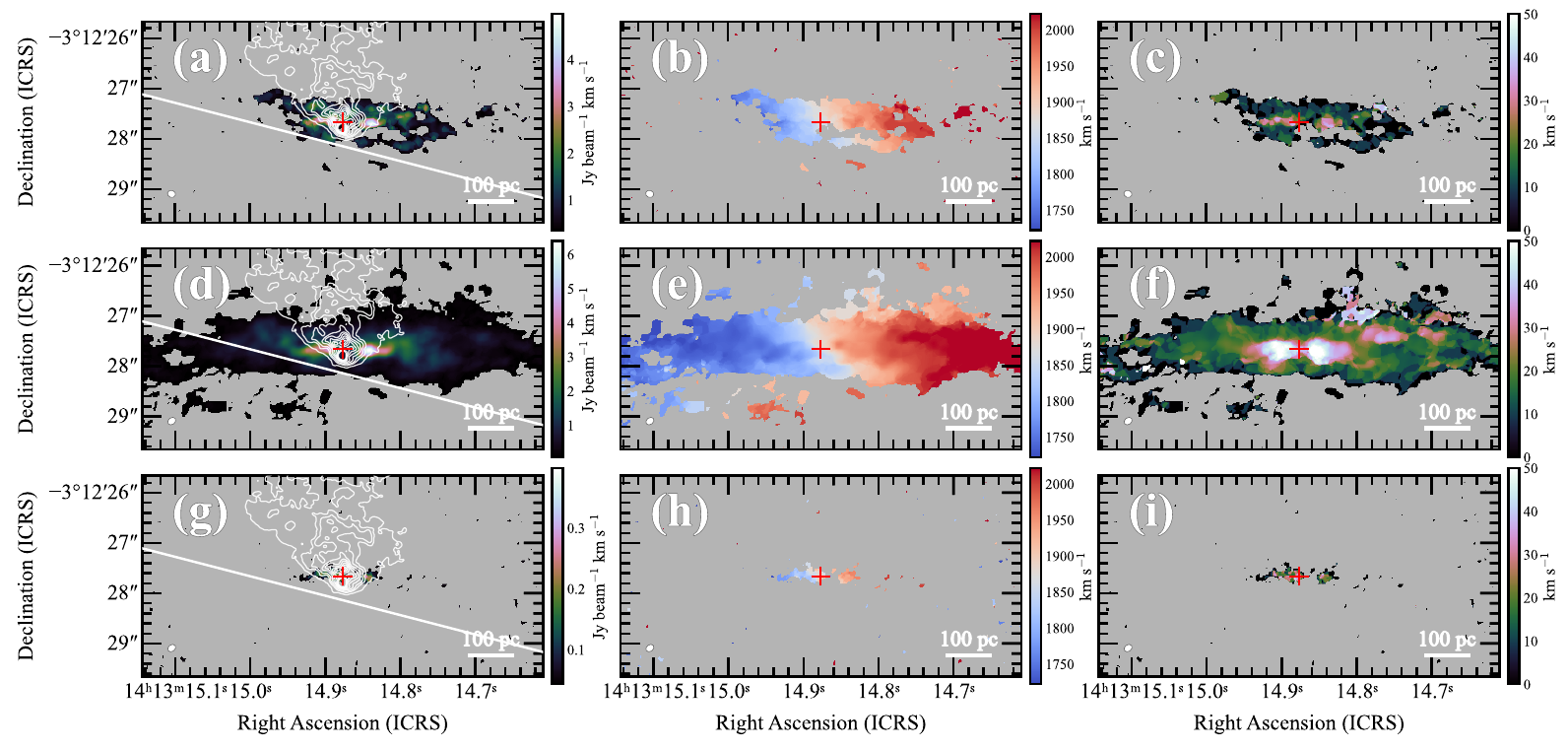}
\caption{Moment maps of (a)--(c) [\ion{C}{1}](1--0), (d)--(f) CO(3--2), and (g)--(i) HCO$^{+}$(4--3). From left to right, panels show the moment 0 maps (integrated intensity; (Jy~beam$^{-1}$~km~s$^{-1}$)), moment 1 maps (line-of-sight velocity; (km~s$^{-1}$)), and moment 2 maps (line-of-sight velocity dispersion; (km~s$^{-1}$)). The red plus sign in each panel marks the AGN position, determined from the peak of the Band 7 continuum emission. 
Contours in the left column show the [\ion{O}{3}]$\lambda5007$ emission at 1$\sigma$, 2$\sigma$, 3$\sigma$, 5$\sigma$, 7$\sigma$, 9$\sigma$, 11$\sigma$, and 13$\sigma$. The white line in the left column indicates the edge of the HST field of view. 
The horizontal and vertical axes represent R.A. and decl. respectively. Moment maps were generated using pixels with signal-to-noise ratios $>3.5\sigma$, where 1$\sigma$ corresponds to 7.7, 2.2, and 0.69~mJy~beam$^{-1}$ for [\ion{C}{1}](1--0), CO(3--2), and HCO$^{+}$(4--3), respectively.} 
\label{fig:momentmaps}
\end{center}
\end{figure*}

\begin{figure}[ht!]
\begin{center}
\includegraphics[width=\linewidth]{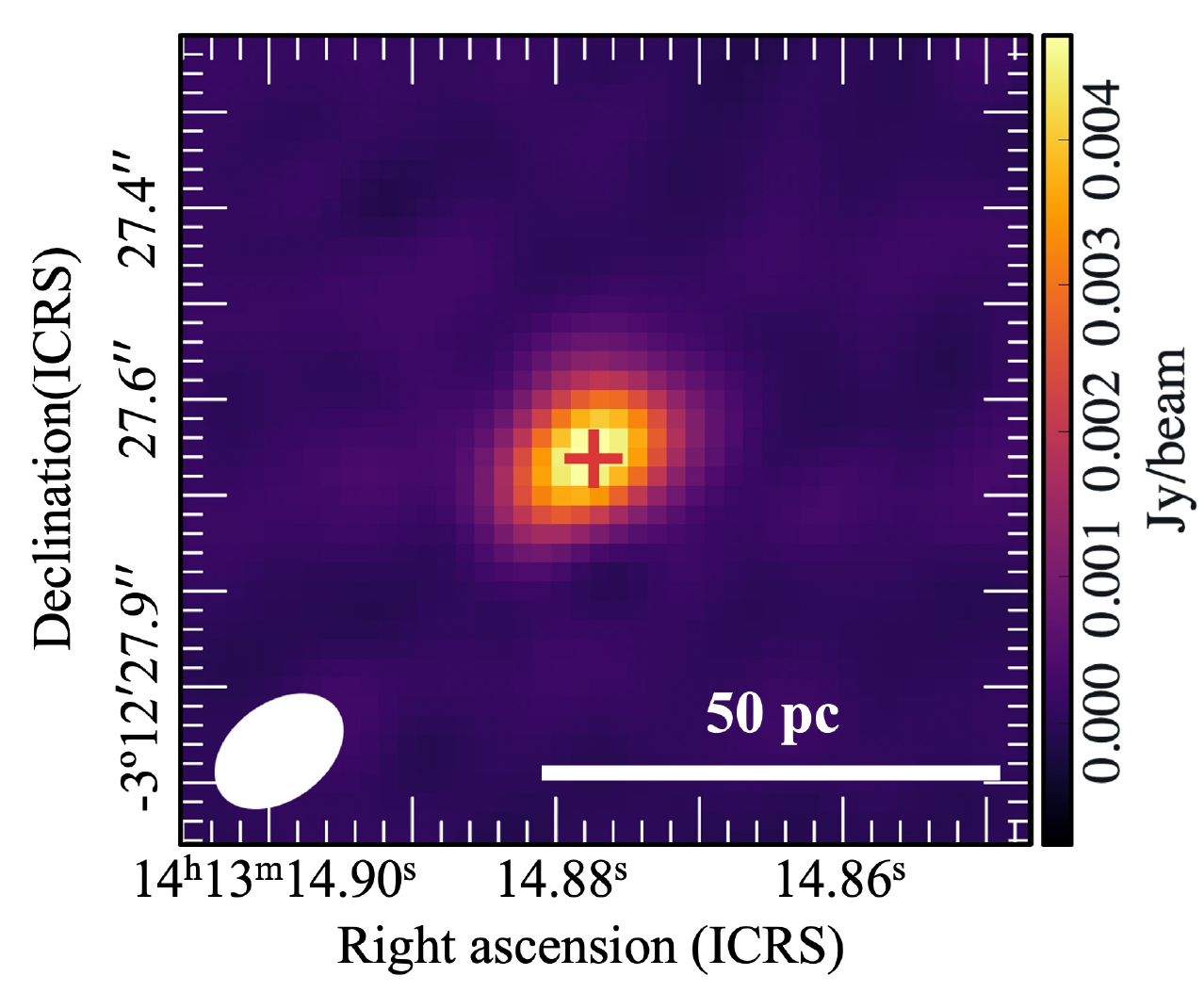}
\caption{ALMA Band 7 continuum image. The red plus sign marks the continuum peak at (R.A., decl.) = (14$^{\rm h}$13$^{\rm m}$14$^{\rm s}$.877, $-$3$\arcdeg$12$\arcmin$27$\arcsec$.66). The white ellipse indicates the synthesized beam. }
\label{fig:cont}
\end{center}
\end{figure}

\begin{figure}
    \centering
    \includegraphics[width=\linewidth]{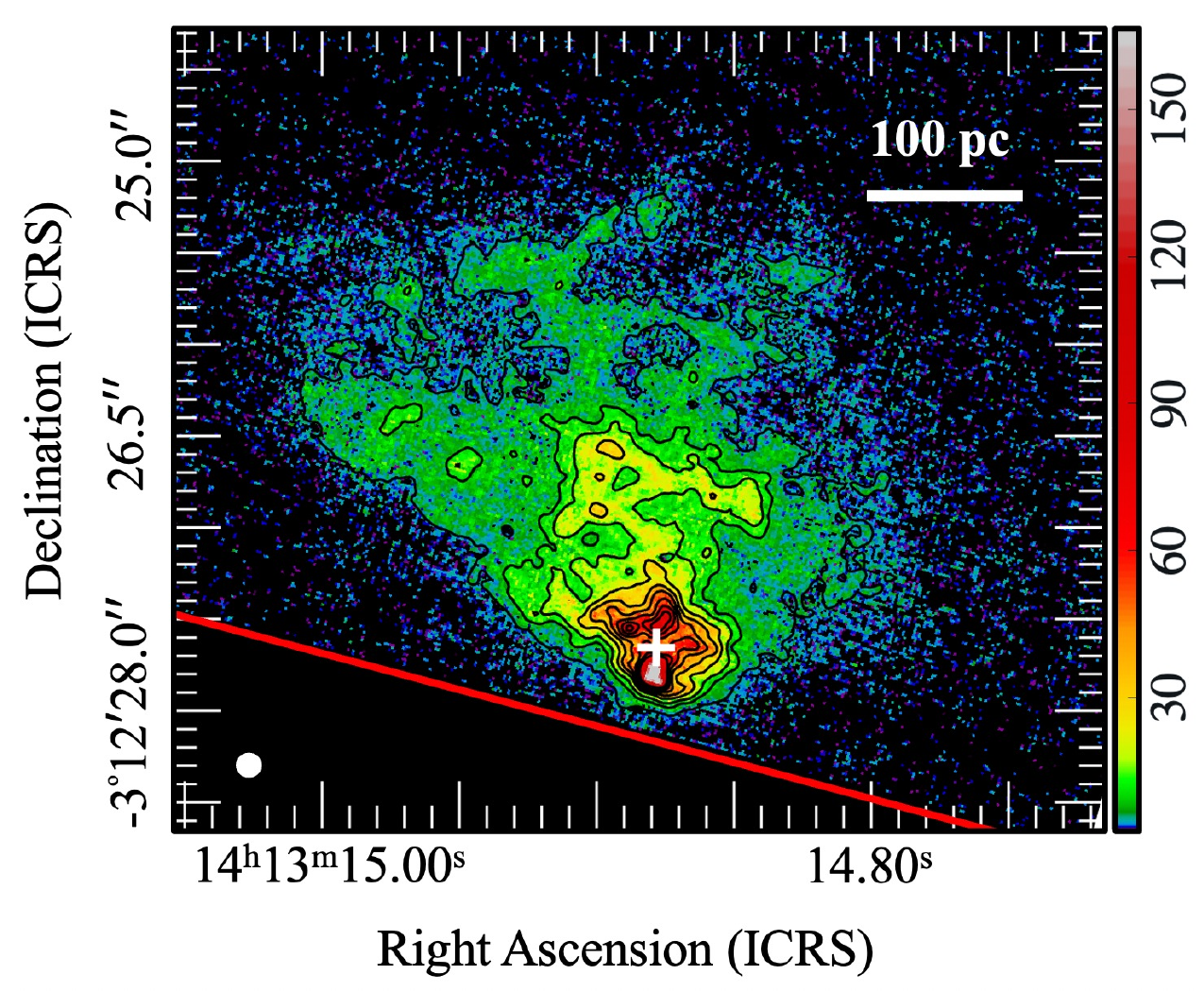}
    \caption{HST/FOC F501N image of NGC~5506. Black contours are shown at 1$\sigma$, 2$\sigma$, 3$\sigma$, 5$\sigma$, 7$\sigma$, 9$\sigma$, 11$\sigma$, and 13$\sigma$. The white plus sign marks the AGN position, and the white circle in the bottom-left corner indicates the point spread function (PSF; FWHM $=0^{\prime\prime}.054$). The red line marks the edge of the HST field of view.}
    \label{fig:OIII}
\end{figure}

\begin{figure*}[ht!]
\begin{center}
\includegraphics[width=\linewidth]{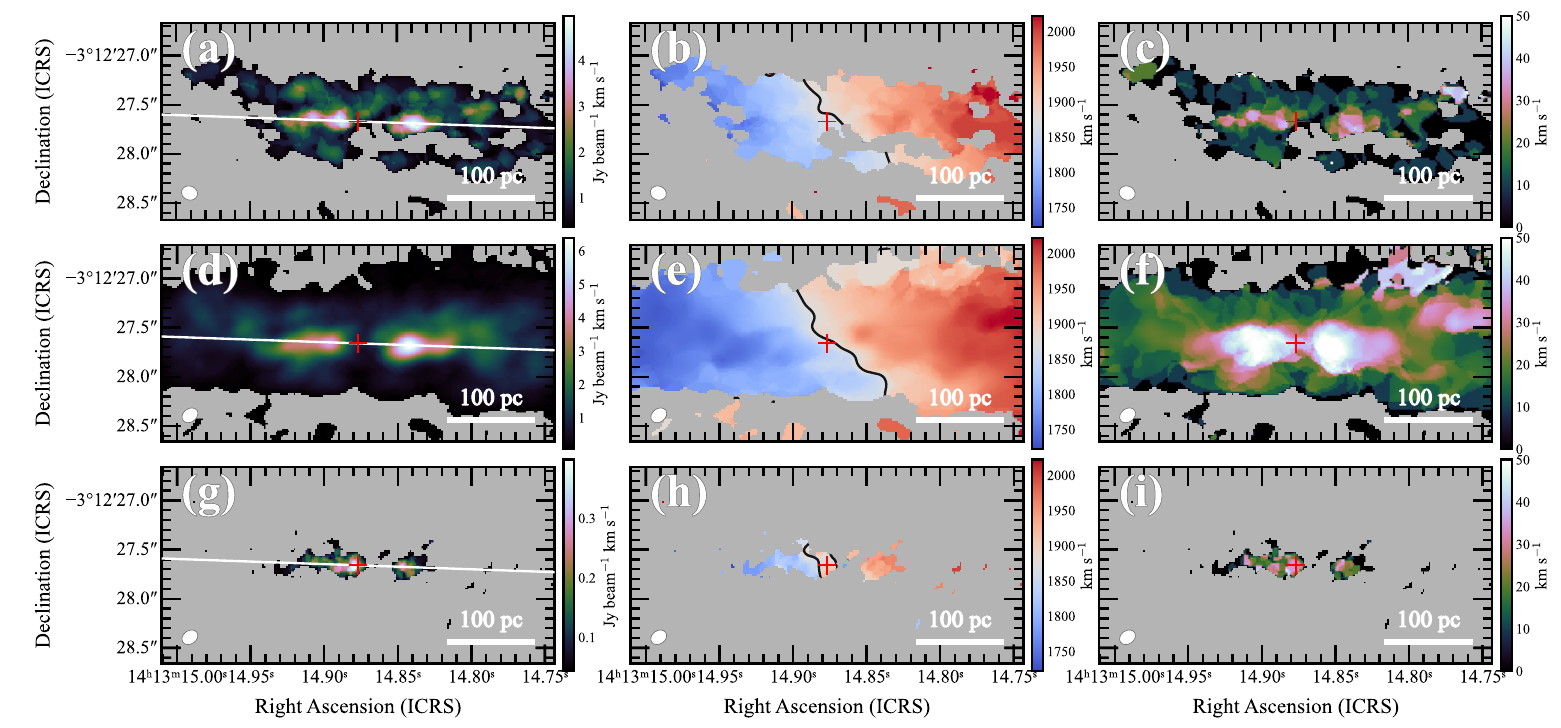}
\caption{Zoomed-in view of the moment maps shown in Figure~\ref{fig:momentmaps}. The white line in panels (a), (d), and (g) indicates the major axis of the CND (position angle = 268$^{\circ}$). The black contour in panels (b), (e), and (h) marks the systemic velocity ($V_{\rm sys}=1872$~km~s$^{-1}$).}
\label{fig:mom0}
\end{center}
\end{figure*}

\begin{figure}[ht!]
\begin{center}
\includegraphics[width=0.8\linewidth]{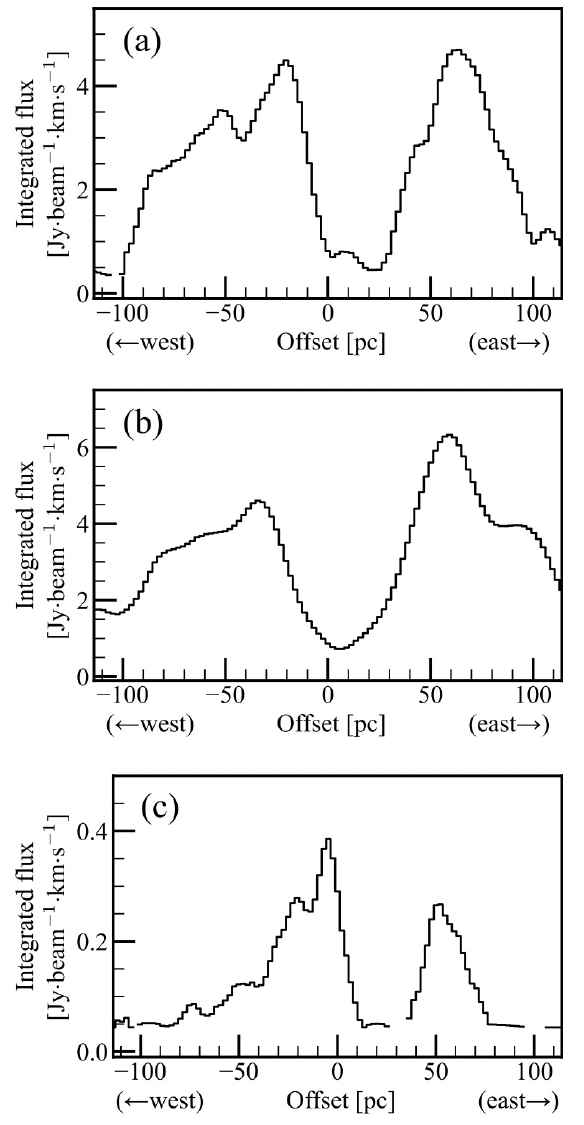}
\caption{Intensity profiles along the major axis of the CND of (a) [\ion{C}{1}](1--0), (b) CO(3--2), and (c) HCO$^{+}$(4--3), extracted from the moment 0 maps over a range of $\pm1^{\prime\prime}$ ($=\pm114$~pc).} 
\label{fig:majslice}
\end{center}
\end{figure}

Figure~\ref{fig:momentmaps} presents the moment maps of [\ion{C}{1}](1--0), CO(3--2), and HCO$^{+}$(4--3), generated using pixels with a signal-to-noise ratio $>3.5$. The CND is defined as the region with emission detected at $>3.5\sigma$ within a 5$^{\prime\prime}$$\times$1$^{\prime\prime}.4$ box centered on the AGN. 
The red plus sign in each panel marks the AGN position (R.A., decl.) = (14$^{\rm h}$13$^{\rm m}$14$^{\rm s}$.877, $-$3$\arcdeg$12$\arcmin$27$\arcsec$.66). This position is determined from the peak of the Band 7 continuum (Figure \ref{fig:cont}), and is consistent with the Band 8 continuum peak. 
Figure~\ref{fig:OIII} shows the HST/FOC F501N image tracing [\ion{O}{3}]$\lambda$5007 emission, with contours at 1$\sigma$, 2$\sigma$, 3$\sigma$, 5$\sigma$, 7$\sigma$, 9$\sigma$, 11$\sigma$, and 13$\sigma$. The red line marks the edge of the HST field of view.

\subsection{Gas Morphology}
\label{sec:resmor}
Figure~\ref{fig:mom0}(a), (d), and (g) show zoomed-in moment 0 maps of [\ion{C}{1}](1--0), CO(3--2), and HCO$^{+}$(4--3). 
The white line in each panel indicates the major axis of the CND (position angle (PA)$=268^{\circ}$), determined from kinematic modeling with 3D-Based Analysis of Rotating Objects via Line Observations~(\texttt{3D Barolo}; see Section \ref{subsec:vdisp}). 

Figure \ref{fig:majslice} shows the intensity profiles along the major axis extracted from the moment 0 maps of [\ion{C}{1}](1--0), CO(3--2), and HCO$^{+}$(4--3).
All tracers exhibit two primary peaks to the east and west of the AGN, at $21\pm9$~pc and $64\pm9$~pc for [\ion{C}{1}](1--0), $35\pm9$~pc and $60\pm9$~pc for CO(3--2), and $5\pm9$~pc and $51\pm9$~pc for HCO$^{+}$(4--3). 
No emission peak is detected at the AGN position in [\ion{C}{1}](1--0) or CO(3--2), whereas HCO$^{+}$(4--3) shows a peak within one beam of the AGN. In all tracers, the midpoint between the two peaks is offset to the west of the AGN.

\subsection {Velocity structure}
The moment 1 maps (Figure \ref{fig:mom0}(b), (e), and (h)) show the flux-weighted mean line-of-sight velocity. 
The systemic velocity ($V_{\rm sys}$) of the CND was determined by fitting a rotating disk model to CO(3--2) data ($1872\pm10$~km~s$^{-1}$; see Section \ref{subsec:vdisp}), and adopted for [\ion{C}{1}](1--0) and HCO$^{+}$(4--3) as well.
In all three tracers, the western side is redshifted, and the eastern side is blueshifted, consistent with rotational motion. 
At the AGN position, the flux-weighted mean line-of-sight velocities are $1859\pm10$~km~s$^{-1}$, $1865\pm10$~km~s$^{-1}$, and $1889\pm10$~km~s$^{-1}$ for [\ion{C}{1}](1--0), CO(3--2), and HCO$^{+}$(4--3), respectively.
The velocity of HCO$^{+}$(4--3) at the AGN position is redshifted by $\sim +17$~km~s$^{-1}$ relative to $V_{\rm sys}$, which may indicate inflow motion toward the AGN. The systemic velocity contours of [\ion{C}{1}](1--0) and CO(3--2) in the moment 1 maps (black lines in Figure \ref{fig:mom0}(b) and (e)) deviate from the kinematic minor axis.

\begin{figure*}[ht!]
\begin{center}
\includegraphics[width=0.7\linewidth]{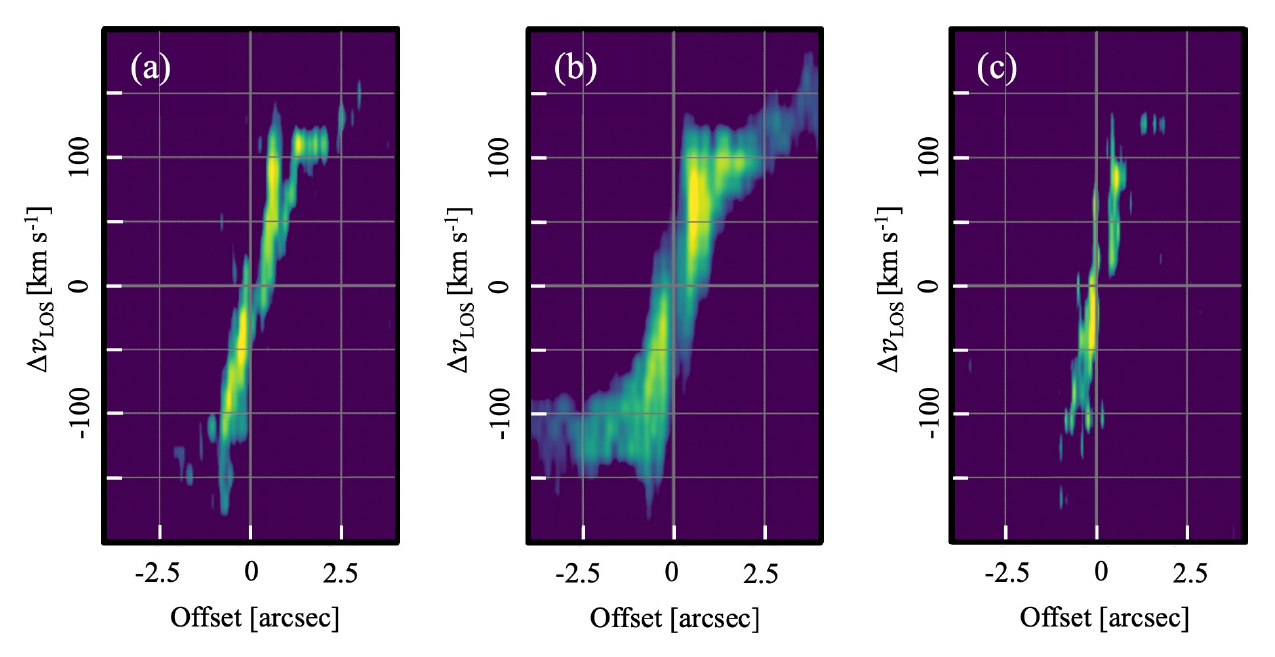}
\caption{Position--velocity diagrams (PVDs) along the major axis of the CND for (a)~[\ion{C}{1}](1--0), (b)~CO(3--2), and (c)~HCO$^{+}$(4--3). The horizontal axis shows the projected distance from the AGN along the major axis, and the vertical axis shows the line-of-sight velocity relative to the systemic velocity.  \label{fig:pvd}}
\end{center}
\end{figure*}

Figure \ref{fig:pvd} shows the position--velocity diagrams (PVDs) extracted along the major axis of the CND. 
The PVDs appear symmetric about both the systemic velocity and the AGN position, confirming the validity of these reference values.
No signatures of a bar structure are evident in the PVDs.

\section{Discussion}\label{sec:discussion}
\subsection{Geometrical Structure}
\subsubsection{Overall Structure}
As described in Section~\ref{sec:resmor}, the moment 0 maps of [\ion{C}{1}](1--0), CO(3--2), and HCO$^{+}$(4--3) each exhibit two intensity peaks, consistent with a torus-shaped CND. Similar off-nuclear double-peaked CO emission has been reported in other highly inclined galaxies, including NGC~7314~\citep{Garcia2021} and NGC~1380~\citep{Kabasares2022}. In contrast, [\ion{O}{3}]$\lambda$5007 shows a conical morphology confirmed spectroscopically by~\citet[][see Section \ref{sec:lineratio} for details]{Fischer2013}.

\subsubsection{Geometrical Thickness of the Circumnuclear Disk}
\label{subsec:geom}

\begin{figure*}[ht!]
\begin{center}
\includegraphics[width=\linewidth]{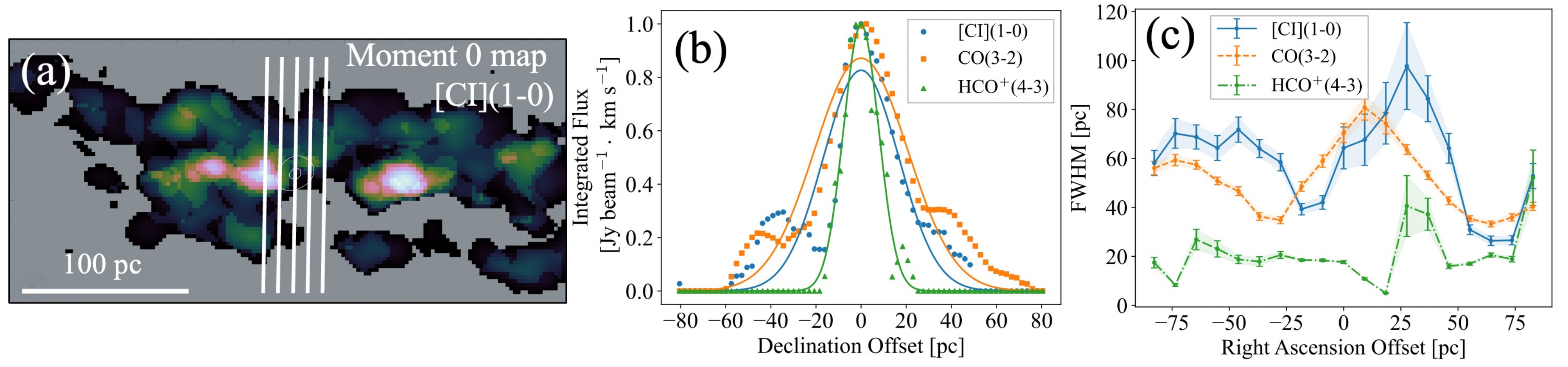}

\caption{Procedure for comparing the geometrical thickness among different gas phases: (a) example lines along which slice profiles are extracted, (b) an example slice profile fitted with a Gaussian, and (c) FWHM of the fitted Gaussians as a function of major-axis position. In panel~(c), the blue solid, orange dashed, and green dashed-dotted lines represent [\ion{C}{1}](1--0), CO(3--2), and HCO$^{+}$(4--3), respectively. The moment 0 map in panel (a) is identical to that shown in Figure~\ref{fig:momentmaps}(a). \label{fig:slice}}
\end{center}
\end{figure*}

We compare the geometrical thickness of the CND among different gas phases.
Because NGC~5506 is highly inclined (inclination $i=79^{\circ}$; see Section \ref{subsec:vdisp}), the geometrical thickness of the CND can be inferred from the emission width along the minor axis direction~\footnote{Figure~2 of \citet{Baba2024} shows that simulations based on the radiation-driven fountain model at $i=80^{\circ}$ predict [\ion{C}{1}] to be geometrically thicker than CO and HCO$^{+}$. If NGC~5506 ($i=79^{\circ}$) has a similar structure, differences in geometrical thickness among the multiphase gas components are expected to be observable.}. 
Slice profiles of the moment 0 maps are extracted along lines parallel to the minor axis (e.g., Figure \ref{fig:slice}(a)) at 4 pixel ($\sim9$~pc) intervals along the major axis and averaged over a 3 pixel ($\sim7$~pc) width. Each profile is fitted with a Gaussian (solid lines in Figure~\ref{fig:slice}(b)), with the peak fixed to the corresponding major axis position. The FWHM of each best-fit Gaussian is then derived as a measure of the emission width. 

Figure~\ref{fig:slice}(c) compares the FWHM values of the three tracers from $-90$ to 90~pc. HCO$^{+}$(4--3) exhibits the smallest FWHM at nearly all positions (18 out of 19). Within 20~pc of the AGN, CO(3--2) shows a marginally larger FWHM than [\ion{C}{1}](1--0), although most differences are within the uncertainties. At larger radii, [\ion{C}{1}](1--0) tends to have a larger FWHM, except in the western region beyond 50~pc of the AGN, where the trend reverses. Throughout the CND, the FWHM values of [\ion{C}{1}](1--0) and CO(3--2) differ by less than a factor of 2 without a systematic trend. We therefore conclude that these two tracers have comparable geometrical thicknesses.
\commentr{Both [\ion{C}{1}](1--0) and CO(3--2) reach maximum FWHM values of $\gtrsim80$~pc, whereas HCO$^{+}$(4--3) forms a thinner disk with a maximum FWHM of $\sim50$~pc. }

\subsection{Gas Dynamics}
\subsubsection{Disk Thickness Inferred from Gas Dynamics}
\label{subsec:vdisp}

\begin{figure*}[ht!]
\begin{center}
\includegraphics[width=\linewidth]{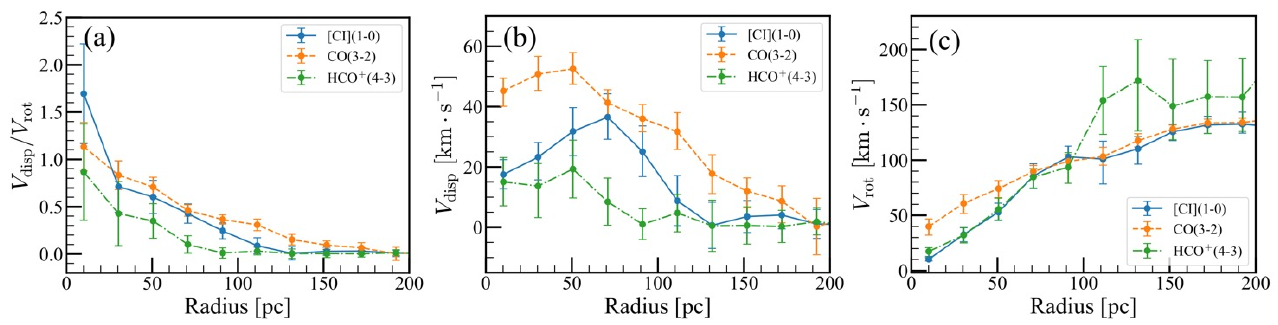}
\caption{(a) Ratio of velocity dispersion to the rotational velocity ($V_{\rm {disp}}$/$V_{\rm {rot}}$), (b) velocity dispersion ($V_{\rm {disp}}$), and (c) rotational velocity ($V_{\rm {rot}}$), derived from \texttt{3D BAROLO} rotating-disk modeling. The blue solid, orange dashed, and green dashed-dotted lines represent [\ion{C}{1}](1--0), CO(3--2), and HCO$^{+}$(4--3), respectively. \label{fig:disptorot}}
\end{center}
\end{figure*}

We model the kinematics of the CND in NGC~5506 using a three-dimensional tilted-ring approach with \texttt{3D BAROLO}~\citep{DiTeodoro&Fraternali2015}. The systemic velocity ($V_{\rm sys}$), inclination ($i$), and PA are first determined from the CO(3--2) data cube, which has the highest signal-to-noise ratio among the three lines. During this step, the radial velocity is fixed to $0$~km~s$^{-1}$. The fit is restricted to radii smaller than 200~pc to focus on the AGN vicinity and avoid contamination from a redshifted region southeast of the AGN that appears unrelated to disk rotation. The ring separation is set to 0.176$^{\prime\prime}$, comparable to the major-axis FWHM of the CO(3--2) synthesized beam. The innermost radius is fixed at 0.088$^{\prime\prime}$, ensuring that the first ring lies outside a single beam element. A single-plane disk model is adopted with warping disabled, as no evidence for a warped disk has been reported in NGC~5506. 
The fit yields $V_{\rm sys}$~=~1872~km~s$^{-1}$, $i=79^{\circ}$, and PA~$=268^{\circ}$, consistent with the values reported by \citet{Esposito2024} ($V_{\rm sys}=1872$~km~s$^{-1}$, $i=80^{\circ}$, and PA$=265^{\circ}$). With $V_{\rm sys}$, $i$, and PA fixed to these values, \texttt{3D BAROLO} is then applied independently to the [\ion{C}{1}](1--0), CO(3--2), and HCO$^{+}$(4--3) data cubes. At each radius, the rotational velocity ($V_{\rm rot}$) and the velocity dispersion ($V_{\rm disp}$) are treated as free parameters. 

The ratio $V_{\rm {disp}}$/$V_{\rm {rot}}$ is plotted as a function of radius in Figure~\ref{fig:disptorot}(a). Under the assumption of hydrostatic equilibrium, \commentr{this ratio serves as a proxy for the disk scale height-to-radius ratio ($H/R$; \citealt{Izumi2018}).}
HCO$^{+}$(4--3) exhibits lower $V_{\rm {disp}}$/$V_{\rm {rot}}$ than the other two tracers at most radii, although the differences are largely within the uncertainties. Within 100~pc, no significant difference in $V_{\rm {disp}}$/$V_{\rm {rot}}$ is found between [\ion{C}{1}](1--0) and CO(3--2). At radii beyond 100~pc, CO(3--2) tends to show higher $V_{\rm {disp}}$/$V_{\rm {rot}}$ than [\ion{C}{1}](1--0) and HCO$^{+}$(4--3). \commentr{The ratio reaches particularly high values ($\gtrsim0.9$) at the innermost radius for both [\ion{C}{1}](1--0) and CO(3--2) (see Section \ref{sec:interp} for further discussion)}, suggesting a high $H/R$. The large uncertainties in $V_{\rm {disp}}$/$V_{\rm {rot}}$ of [\ion{C}{1}](1--0) and HCO$^{+}$(4--3) at the innermost radius arise from the small rotational velocity at that radius. 
The comparable $V_{\rm {disp}}$/$V_{\rm {rot}}$ of [\ion{C}{1}](1--0) and CO(3--2) and the lower ratio of HCO$^{+}$(4--3) within 100~pc are consistent with the geometrical thickness comparison presented in Section~\ref{subsec:geom}.

\subsubsection{Radial Velocity}
\label{subsec:barolo}

\begin{figure*}[ht!]
\begin{center}
\includegraphics[width=0.8\linewidth]{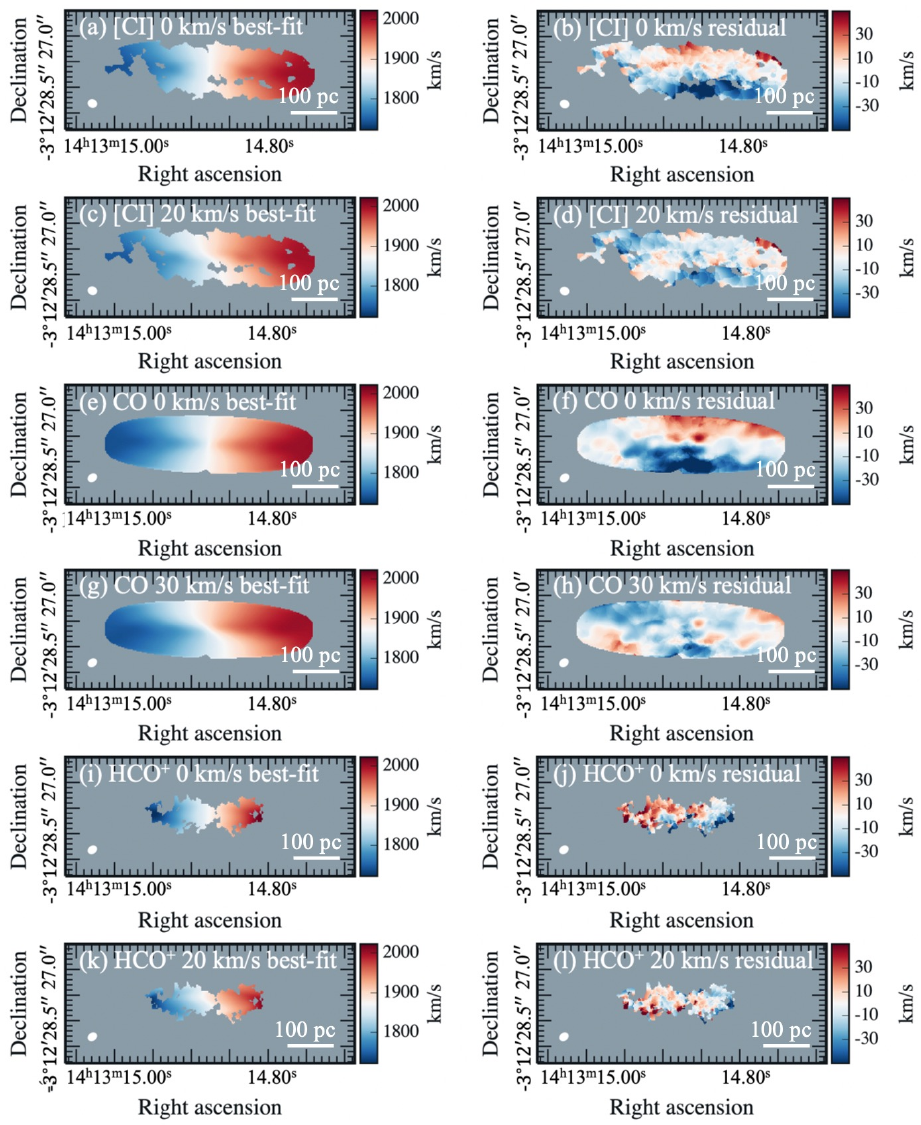}

\caption{(a), (e), and (i): best-fit moment 1 maps of [\ion{C}{1}](1--0), CO(3--2), and HCO$^{+}$(4--3) obtained without outflow velocity. (b), (f), and (j): residual maps corresponding to panels (a), (e), and (i). (c), (g), and (k): best-fit moment 1 maps of [\ion{C}{1}](1--0), CO(3--2), and HCO$^{+}$(4--3) obtained with the optimal outflow velocity included. (d), (h), and (l): residual maps corresponding to panels (c), (g), and (k). 
\label{fig:outflows}
}
\end{center}
\end{figure*}

We derive the mean radial velocities of [\ion{C}{1}](1--0), CO(3--2), and HCO$^{+}$(4--3). Figure~\ref{fig:outflows}(a), (e), and (i) show the best-fit moment~1 maps without an outflow component. 
The corresponding residual maps (Figure~\ref{fig:outflows}(b), (f), and (j)) exhibit a systematic redshift on the northern side and blueshift on the southern side, consistent with a radial outflow. We therefore include a constant outward radial velocity (hereafter outflow velocity, $V_{\rm out}$) and repeat the fitting with $V_{\rm out}$ values from 0 to 40~km~s$^{-1}$ in steps of 10~km~s$^{-1}$.
The best-fit models, including the optimal outflow that minimizes the residual rms are shown in Figure~\ref{fig:outflows}(c), (g), and (k). 
The optimal outflow velocities are 20$~\pm~5$~km~s$^{-1}$ for [\ion{C}{1}](1--0), 30$~\pm~5$~km~s$^{-1}$ for CO(3--2), and 20$~\pm~5$~km~s$^{-1}$ for HCO$^{+}$(4--3). The optimal outflow velocity of CO(3--2) is consistent with the value reported by \citet[][26$~\pm~$9~km~s$^{-1}$)]{Esposito2024}.
Because CO(3--2) has the largest spatial extent among the three tracers, it samples emission from outer radii where the outflow velocity is higher (Figure \ref{fig:outflows}(f)). This likely accounts for the higher $V_{\rm out}$ inferred from CO(3--2) relative to [\ion{C}{1}](1--0) and HCO$^{+}$(4--3).

\subsection{[\ion{C}{1}](1--0)/CO(3--2) Line Ratio}
\label{sec:lineratio}

\begin{figure*}[ht!]
\begin{center}
\includegraphics[width=0.8\linewidth]{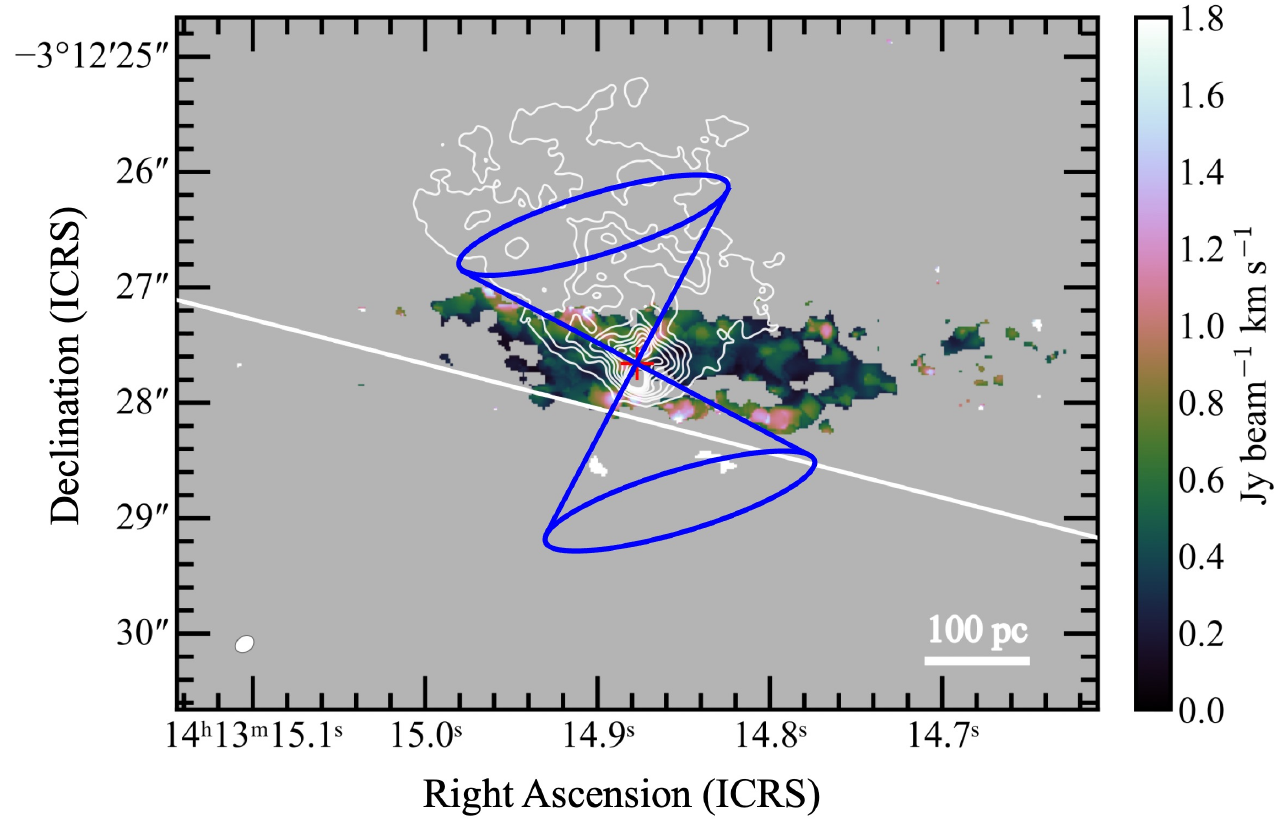}
\caption{[\ion{C}{1}](1--0) to CO(3--2) integrated intensity ratio ($R_{\rm [C~I]/CO}$) map on a brightness temperature scale, overlaid with contours of [\ion{O}{3}]$\lambda$5007 emission (white; 1$\sigma$, 2$\sigma$, 3$\sigma$, 5$\sigma$, 7$\sigma$, 9$\sigma$, 11$\sigma$, and 13$\sigma$) and the corresponding biconical outflow model~\citep[blue;][]{Fischer2013}. The red cross marks the AGN position. \label{fig:lineratoiii}}
\end{center}
\end{figure*}

We derive [\ion{C}{1}](1--0)/CO(3--2) integrated intensity ratio ($R_{\rm [C~I]/CO}$) from the moment 0 maps. 
Because atomic carbon can be produced through CO dissociation by high-energy photons, this ratio is sensitive to the local radiation field and thus provides insight into the origin of the multiphase gas. 

The color map in Figure~\ref{fig:lineratoiii} shows the spatial distribution of $R_{\rm [C~I]/CO}$ on a brightness temperature scale. 
The mean ratio across the entire CND is $0.52\pm0.07$, higher than values typical of star-forming galaxies~\citep[$\sim 0.1$;][]{Liu2023}. This suggests that high $R_{\rm [C~I]/CO}$ is driven by nonstellar processes, such as X-ray irradiation from the AGN or interactions between the CND and an AGN-driven biconical ionized outflow.
Additionally, regions to the northeast and southwest of the AGN exhibit particularly elevated ratios. 
Although AGN X-ray emission likely contributes to CO dissociation across the CND, it cannot account for the nonaxisymmetric distribution of the elevated $R_{\rm [C~I]/CO}$ regions. We therefore focus on the outflow interaction scenario in this section.

\begin{table}[ht!]
\centering 
\caption{Integrated Line Ratio (\commentbf{$R_{\rm [C~I]/CO}$})}
\begin{tabular}{c c c c} 
\hline\hline 
 & \commentbf{$R_{\rm [C~I]/CO}$}\\ 
\hline 
CND &  $0.52\pm0.07$\\ 
Inside the bicone & $0.82\pm0.11$ \\
Outside the bicone & $0.47\pm0.07$ \\
\hline\hline 
\end{tabular}
\label{tbl:linerat} 
\tablecomments{The $10\%$ absolute flux uncertainties are included.}
\end{table}

The $R_{\rm [C~I]/CO}$ map in Figure~\ref{fig:lineratoiii} is overlaid with contours of [\ion{O}{3}]$\lambda$5007 emission (white) and the corresponding biconical outflow model~\citep[blue line;][]{Fischer2013}. The figure shows a spatial correlation between the [\ion{O}{3}]$\lambda$5007 emission and the elevated $R_{\rm [C~I]/CO}$ regions.
Ionized gas on the southern side of the AGN has been detected in [\ion{O}{3}]~\citep{Esposito2024} and [\ion{Ne}{5}] \citep{Zhang2024}, but the southern ionization cone is not detected in the HST observations presented here for the following reasons. 
The CND has an inclination of $79^{\circ}$, with the northern face tilted 11$^\circ$ from edge-on toward the observer. The southern cone thus lies on the far side of the disk and is projected against the foreground molecular disk traced by CO(3--2) (Figure \ref{fig:momentmaps}(d)), where line-of-sight obscuration by molecular gas and/or dust likely suppresses the [\ion{O}{3}] emission. Furthermore, the less obscured regions of the southern cone are outside the field of view of this observation.

Table~\ref{tbl:linerat} summarizes the mean $R_{\rm [C~I]/CO}$ for the entire CND and for regions inside and outside the bicone. The mean ratio inside the bicone ($0.82\pm0.11$) is higher than that outside the bicone ($0.47\pm0.07$). This suggests that CO is preferentially dissociated at the interface where the ionized outflow interacts with the CND. A plausible mechanism is that the outflow drives shocks into the molecular gas, and the resulting shock heating and/or shock-accelerated cosmic rays enhance CO dissociation. 
A similar trend has been reported in another Seyfert 2 galaxy, NGC~1068. The [\ion{C}{1}](1--0)/CO(1--0) intensity ratio in NGC~1068 is spatially correlated with the [\ion{S}{3}]/[\ion{S}{2}], a tracer of the ionization cone~\citep{Saito2022}. \citet{Saito2022} suggest that this correlation arises from the dissociation of CO into atomic carbon by the AGN-driven biconical ionized outflow, consistent with the scenario proposed here for NGC~5506.

\subsection{Interpretation of Morphology and Kinematics of [\ion{C}{1}](1--0) and CO(3--2) Lines}
\label{sec:interp}
In this section, we discuss the factors contributing to the geometrical thickness of the CND. As shown in Figure~\ref{fig:slice}(c), [\ion{C}{1}](1--0) and CO(3--2) show no significant differences in geometrical thickness, whereas HCO$^{+}$(4--3) is more concentrated toward the disk plane.
These results are supported by the $V_{\rm disp}/V_{\rm rot}$ ratio\commentr{, a proxy for $H/R$,} shown in Figure~\ref{fig:disptorot}(a). Within a radius of 100~pc, this ratio is comparable between [\ion{C}{1}](1--0) and CO(3--2). The ratio reaches particularly high values (\commentr{$\gtrsim0.9$}) in the central region ($\lesssim30$~pc), higher than the value reported for the Circinus galaxy~\commentr{\citep[$\lesssim0.5$;][]{Izumi2018}}.
High $H/R$ values in molecular gas have also been reported in other Seyfert galaxies based on the H$_2(\nu=$1--0 S(1), rest frame 2.1~$\mu$m) line~\citep[][see also \citealt{Sani2012} for $H/R$ of dense molecular gas]{Hicks2009}. 

The radiation-driven fountain model~\citep{Wada2012} proposes that a geometrically thick torus-shaped structure can be supported through gas circulation driven by AGN radiative feedback. \citet{Schartmann2014} demonstrate that such fountain flows reproduce the observed differences in the spectral energy distributions (SEDs) of type-1 and type-2 Seyfert galaxies. Additionally, the polar dust emission that dominates the mid-infrared range \citep{Asmus2016} is naturally explained within this framework.
\citet{Wada2016} extend the radiation-driven fountain model to incorporate SN feedback and nonequilibrium chemistry in X-ray-dominated regions. Their simulations show that atomic hydrogen forms a geometrically thick torus, molecular hydrogen (H$_2$) is confined primarily to the disk plane, and ionized gas forms a biconical structure. Simulations by \citet{Baba2024} further predict distinct spatial distributions for C, CO, and HCO$^{+}$; neutral carbon forms a geometrically thick torus while CO and HCO$^{+}$ are confined primarily to the disk plane.

\citet{Wada2015} shows that the structure of the fountain flow depends on both the black hole mass and the AGN luminosity (or Eddington ratio); the scale height increases with Eddington ratio and decreases with black hole mass. Once the fountain flow is established, variations in the radiation field and thermal conditions produce distinct spatial distributions of atomic and molecular gas. However, when the Eddington ratio is too low (e.g., $\lambda_{\rm Edd}\sim 0.01$), a stable fountain flow cannot be sustained, and clear differences between the atomic and molecular gas distributions are not expected. 
For NGC~5506, the Eddington ratio $\lambda_{\rm Edd}=0.05^{\rm +0.21}_{\rm -0.04}$ \citep{Esposito2024} is lower than the value adopted in the simulation ($\lambda_{\rm Edd}=0.2$), and the black hole mass $M_{\rm BH}=2.0^{+8.0}_{-1.6}\times 10^7 M_{\odot}$~\citep{Gofford2015} exceeds that assumed in the simulation ($M_{\rm BH}=2.0\times 10^6 M_{\odot}$). 
These differences suggest that the fountain flow in NGC~5506 is likely less prominent than in the simulation, and the structural differences predicted by this model may not be detectable in this system.

\begin{figure*}
    \centering
    \includegraphics[width=0.8\linewidth]{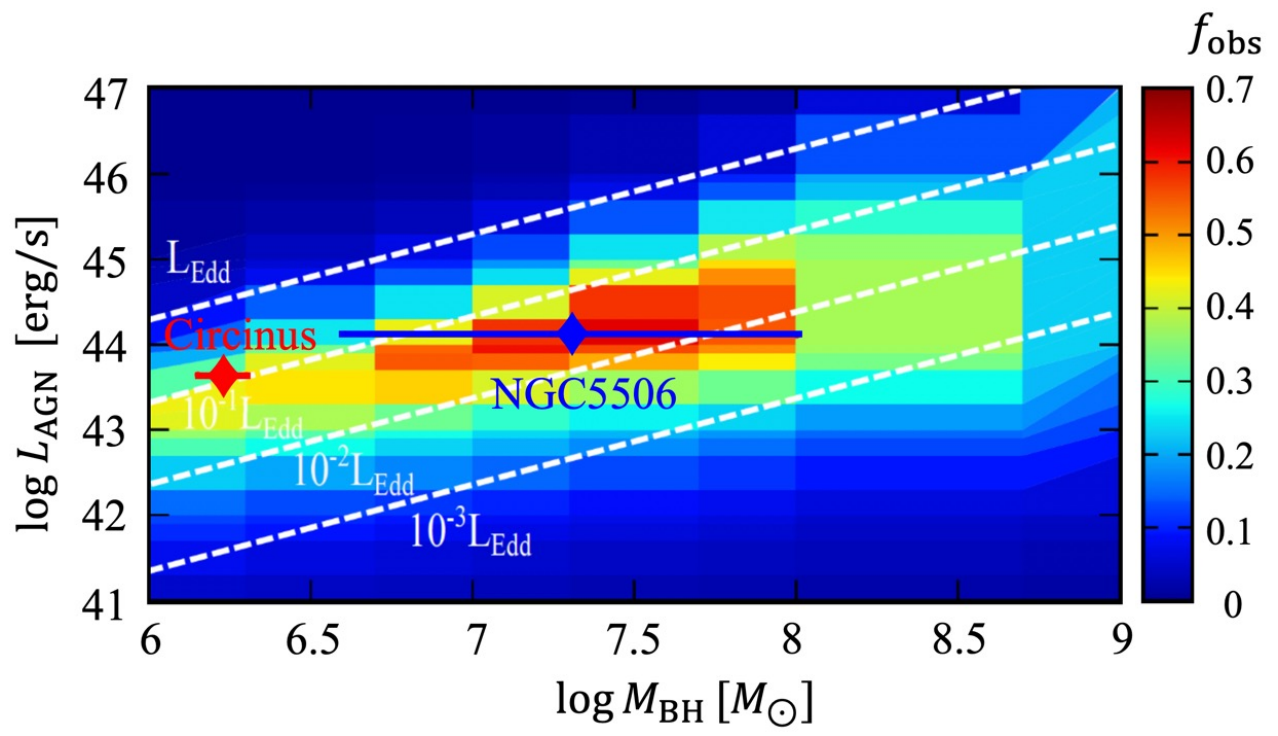}
    \caption{Obscuring fraction ($f_{\rm obs}$) as a function of black hole mass ($M_{\rm BH}~[M_\odot]$) and AGN luminosity ($L_{\rm AGN}$~[erg$~$s$^{-1}$]) from Figure~4 of \citet{Kawakatu2020}, with observational measurements overplotted. \commentr{The obscuring fraction ($f_{\rm obs}$) and scale height ($H/R$) are related by $H/R=f_{\rm obs}$/$\sqrt{(1-f_{\rm obs}^2})$ (see Section~\ref{sec:interp} for details).} The blue and red diamonds with horizontal error bars indicate measured values and their uncertainty for NGC~5506 and the Circinus galaxy, respectively. White dashed contours show constant Eddington ratios of 1, 0.1, 0.01, and 0.001 from top to bottom. \label{fig:fobs}}
\end{figure*}

The high $H/R$ in NGC~5506 is consistent with the SN-driven turbulence model of \citet{Kawakatu2020}, which examines the combined effects of SN and AGN feedback. The SN feedback is based on the SMBH--CND coevolution model of \citet{Kawakatu2008}. 
In their model, SN feedback increases the disk scale height, while AGN feedback expels gas and reduces it. 
\citet{Kawakatu2020} derive the AGN luminosity and black hole mass dependence of the obscuring fraction $f_{\rm obs}$ ($=\cos{\theta_{\rm CND}}$, where $\theta_{\rm CND}$ is the minimum opening angle of the CND), a measure of the disk thickness. As illustrated in Figure \ref{fig:fobs}, $f_{\rm obs}$ reaches a maximum at Eddington ratios of $0.01\text{--}0.1$ and black hole masses of $10^{7\text{--}8}~M_\odot$. 
Within this framework, $f_{\rm obs}$ is estimated to be $\gtrsim0.6$ for NGC~5506, and $\sim0.3$ for the Circinus galaxy. These values ($f_{\rm obs} = 0.3$ and 0.6) correspond to $H/R \approx 0.3$ and $\approx~$0.9, respectively, via the relation $H/R=f_{\rm obs}$/$\sqrt{(1-f_{\rm obs}^2})$. 
\commentr{These theoretical predictions are consistent with the values measured in NGC~5506 in this work and in the Circinus galaxy~\citep{Izumi2018}, indicating that this model can explain the larger scale height of the CND in NGC~5506 relative to the Circinus galaxy.}

Furthermore, the SN-driven turbulent velocity predicted by this model is broadly consistent with the velocity dispersion observed in NGC~5506 at a radius of 100~pc. The turbulent velocity ($v_{\rm t}$) is given by
\begin{equation}
    v_{\rm t}=18~\mathrm{km~s^{-1}}\bigg(\frac{C_\ast}{10^{-8}\mathrm{yr^{-1}}}\bigg)^{1/2}\bigg(\frac{M_{\rm BH}}{10^7M_\odot}\bigg)^{-1/4}\bigg(\frac{r}{30~\mathrm{pc}}\bigg)^{3/4},
\end{equation}
where $C_\ast=\mathrm{star\ formation\ rate\ (SFR)}/M_{\rm gas}$ is the star formation efficiency, $M_{\rm BH}$ is the black hole mass, and $r$ is the distance from the black hole~\citep{Kawakatu2020}. The star formation rate of NGC~5506 is SFR$=0.3^{+0.7}_{-0.2}~M_\odot~\mathrm{yr^{-1}}$, derived from the 11.3~$\mu$m polycyclic aromatic hydrocarbon feature~\citep{Ruschel2017}, and the molecular gas mass ($M_{\rm gas}$) is ~$(4.1\pm 0.6)\times10^{7}~M_\odot$ (Appendix~\ref{subsec:coldens}). 
The predicted turbulent velocity is $v_{\rm t}\sim30~\mathrm{km~s^{-1}}$ at $r\sim100$~pc. This is consistent with the observed velocity dispersion of [\ion{C}{1}](1--0) and CO(3--2), $V_{\rm disp}=20\text{--}40~\mathrm{km~s^{-1}}$. 
Because the radial dependence of the velocity dispersion differs between the model and the observation, the agreement between $v_{\rm t}$ and $V_{\rm disp}$ is less robust at radii other than 100~pc. Nevertheless, the order-of-magnitude consistency supports the interpretation that SN-driven turbulence provides the dominant source of vertical support for the CND in NGC~5506.
A similar correspondence between observed velocity dispersion and modeled turbulent velocity has been reported for NGC~1275~\citep{Nagai2021}.

\section{Conclusion} \label{sec:conclusion}
We investigated the multiphase gas morphology and kinematics in the circumnuclear region of the nearby Seyfert galaxy, NGC~5506. We used ALMA observations of [\ion{C}{1}](1--0), CO(3--2), and HCO$^{+}$(4--3) at a spatial resolution of $\sim$20~pc and HST observations of [\ion{O}{3}]$\lambda$5007. The main findings are summarized as follows:

\begin{enumerate}

    \item{[\ion{C}{1}](1--0), CO(3--2), and HCO$^{+}$(4--3) trace circumnuclear disk structures on scales of several hundred parsecs, while [\ion{O}{3}]$\lambda$5007 exhibits a conical morphology. 
    The geometrical thicknesses of [\ion{C}{1}](1--0) and CO(3--2) are comparable, whereas HCO$^{+}$(4--3) is more concentrated toward the disk plane. }

    \item{The velocity structures of [\ion{C}{1}](1--0) and CO(3--2) are similar. Three-dimensional rotating disk modeling with \texttt{3D BAROLO} reveals comparable velocity dispersion to rotational velocity ratios ($V_{\rm {disp}}$/$V_{\rm {rot}}$, a proxy for disk scale height-to-radius ratio) for [\ion{C}{1}](1--0) and CO(3--2), while HCO$^{+}$(4--3) exhibits systematically lower $V_{\rm {disp}}$/$V_{\rm {rot}}$. In the central region ($\lesssim30$~pc), $V_{\rm {disp}}$/$V_{\rm {rot}}$ is high ($\gtrsim0.9$) for both [\ion{C}{1}](1--0) and CO(3--2), indicating geometrically thick structures. 
    These kinematic results are consistent with the direct comparison of geometrical thickness.}

    \item{The [\ion{C}{1}](1--0)/CO(3--2) ratio ($R_{\rm [C~I]/CO}$) is elevated within the AGN-driven biconical ionized outflow traced by [\ion{O}{3}]$\lambda$5007, with mean values of $0.82\pm0.11$ inside the bicone and $0.47\pm0.07$ outside the bicone.
    This spatial correlation suggests that CO is preferentially dissociated at the interface where the ionized outflow interacts with the CND.}

    \item{The observed CND properties are consistent with a scenario in which SN-driven turbulence provides the vertical support for the CND. The relatively low Eddington ratio and high black hole mass of NGC~5506 suggest that a stable radiation-driven fountain flow may not be established. This may explain the absence of significant differences in geometrical thickness between [\ion{C}{1}](1--0) and CO(3--2). Instead, SN feedback can effectively increase the CND scale height for the black hole mass and luminosity of NGC~5506, consistent with the higher $V_{\rm {disp}}$/$V_{\rm {rot}}$ observed here compared to the Circinus galaxy. The order-of-magnitude agreement between the SN-driven turbulent velocity and the observed velocity dispersion further supports this interpretation.}
    
\end{enumerate}

\begin{acknowledgments}
We thank the anonymous referee for constructive comments that greatly improved this manuscript. K.T. thanks Dai Hirashima for valuable discussions. This paper makes use of the following ALMA data: ADS/JAO.ALMA\#2017.1.00082.S, ADS/JAO.ALMA\#2022.1.00410.S. ALMA is a partnership of ESO (representing its member states), NSF (USA) and NINS (Japan), together with NRC (Canada), NSTC and ASIAA (Taiwan), and KASI (Republic of Korea), in cooperation with the Republic of Chile. The Joint ALMA Observatory is operated by ESO, AUI/NRAO and NAOJ. This research is also based on observations made with the NASA/ESA Hubble Space Telescope obtained from the Space Telescope Science Institute, which is operated by the Association of Universities for Research in Astronomy, Inc., under NASA contract NAS 5–26555. These observations are associated with program 5144. {\it HST} data analyzed in this paper can be found in MASTat \dataset[10.17909/j3py-8853]{http://dx.doi.org/10.17909/j3py-8853}. The Pan-STARRS1 Surveys (PS1) and the PS1 public science archive have been made possible through contributions by the Institute for Astronomy, the University of Hawaii, the Pan-STARRS Project Office, the Max-Planck Society and its participating institutes, the Max Planck Institute for Astronomy, Heidelberg and the Max Planck Institute for Extraterrestrial Physics, Garching, The Johns Hopkins University, Durham University, the University of Edinburgh, the Queen's University Belfast, the Harvard-Smithsonian Center for Astrophysics, the Las Cumbres Observatory Global Telescope Network Incorporated, the National Central University of Taiwan, the Space Telescope Science Institute, the National Aeronautics and Space Administration under Grant No. NNX08AR22G issued through the Planetary Science Division of the NASA Science Mission Directorate, the National Science Foundation Grant No. AST-1238877, the University of Maryland, Eotvos Lorand University (ELTE), the Los Alamos National Laboratory, and the Gordon and Betty Moore Foundation. ChatGPT was used to assist with English polishing. This work is supported by the JSPS KAKENHI grant Nos. JP19K03918 (N.K.), JP21H04496 (K.W. and T.I.), JP22H00157 (M.K.), and JP24K07091 (M.S.). 

\end{acknowledgments}

\facilities{ALMA, HST (STIS)}

\begin{contribution}
K.T. analyzed the data, produced figures, and wrote and submitted the manuscript.
H.N. developed the initial research concept and supervised K.T.
\commentbf{M.S. calibrated the HST data. }
All authors contributed to the scientific discussion. 

\end{contribution}

\appendix
\section{Molecular Gas Mass and Column Density}\label{subsec:coldens}

The H$_2$ gas mass and column density are derived from the moment~0 map of CO(3--2), following~\citet{Izumi2018}.
The CO(3--2) line luminosity $L^\prime_{\rm CO(3-2)}$ is given by
\begin{equation}
    L^\prime_{\rm CO(3-2)}=3.25\times10^7S_{\rm CO(3-2)}\Delta v~\nu^{-2}_{\rm obs}D_{\rm L}^2(1+z)^{-3},
    \label{eq:solomon}
\end{equation}
where $S_{\rm CO(3-2)}\Delta v$ is the velocity-integrated flux density in Jy~km~s$^{-1}$, $\nu_{\rm obs}$ is the observed frequency in GHz, $D_{\rm L}$ is the luminosity distance in Mpc, and $z$ is the redshift~\citep{Solomon&Bout2015}. 
Assuming CO(3--2) emission is thermalized with that of CO(2--1)~\citep{Solomon&Bout2015}, we adopt the canonical CO-to-H$_2$ conversion factor for active environments $M_{\rm H_{2}} / L^{\prime}_{\rm CO} = 0.8 M_{\odot}$~(K~km~s$^{-1}$~pc$^2$)$^{-1}$~\citep{Downes&Solomon1998}. The resulting total molecular hydrogen mass is $M_{\rm H_2}=(4.1\pm 0.6)\times10^7$~$M_\odot$, and the mean column density is $N_{H_2}=(4.3 \pm 0.6)\times10^{22}$~cm$^{-2}$.

\end{document}